\documentclass[floats,aps,prb,draft,preprint,showpacs]{revtex4}

\begin{document}
\title{Electronic transport through a contact of a correlated quantum wire with leads of higher dimension}

\author{S.N. Artemenko}\email[E-mail: ]{art@cplire.ru}
\author{P.P. Aseev} \author{D.S. Shapiro}

\affiliation{V.A. Kotel'nikov Institute for Radioengineering and Electronics of
Russian Academy of Sciences,
Mokhovaya str. 11-7, Moscow 125009, Russia}

\date{\today}

\begin{abstract}
We study theoretically electronic transport through a contact of a quantum wire with 2D or 3D leads and find that if the contact is not smooth and adiabatic then the conduction is strongly suppressed below a threshold voltage $V_T$, while above $V_T$ the dc current $\bar I$ is accompanied by coherent oscillations of frequency $f=\bar I / e$. The effect is related to interelectronic repulsion and interaction of dc current with the Friedel oscillations near a sharp contact. In short conducting channels of length $L < L_0 \simeq  \hbar v_F/eV_T$ and at high temperatures $T > T_0 \simeq eV_T/k_B$ the effect is destroyed by fluctuations.
\end{abstract}

\maketitle

In contrast with 2D and 3D systems where basic electronic properties are usually well described in terms of the Fermi liquid and single-electron noninteracting quasipariticles, 1D systems of interacting electrons are better described in terms of the Luttinger liquid (LL) with bosonic excitations. The LL is an alternative to the Fermi liquid in 1D (for a review see Ref.~\onlinecite{Giamarchi}). Interaction in 1D systems greatly affects both the electronic structure and transport. In particular, a power-law suppression of density of states arises near the Fermi energy and even isolated impurities strongly suppress conduction resulting in power-law dependence of conductance on voltage and/or temperature~\cite{Giamarchi}. This behavior was confirmed experimentally in various 1D systems including semiconductor quantum wires~\cite{Auslaender} and carbon nanotubes~\cite{carbon}, and was described in terms of macroscopic tunneling between different minima of a periodic potential tilted by external bias. The periodic potential is associated with the Friedel oscillations (FO) induced by an impurity. It was shown recently~\cite{ARS} that the power-law regime takes place only at small enough voltages, while above a threshold voltage a dynamical regime of conduction sets in. In this regime the dc current is accompanied by oscillations of frequency $f = I/e$. The effect is induced by motion of the FO in a repulsive electronic system when dc current passes an impurity. Interesting effects can also be expected at sharp, non-adiabatic, contacts between quantum wires and electrodes of higher dimension, i.e. between the Fermi liquid and strongly correlated electronic state. This problem has not been studied yet, and usually the boundary conditions derived by Egger and Grabert~\cite{Grabert} for ideal adiabatic contacts are used. These conditions were derived only for expectation values and, therefore, cannot be used to describe fluctuations, which are very important in 1D systems. Further, real contacts are not necessarily adiabatic and one can expect that reflections of electrons from 
contacts may result in the FO and, hence, in effects similar to those predicted for the case of impurities. Here we derive boundary conditions for formation of the FO near contacts and show that the conductance is affected by the FOs, resulting in the dynamic regime of conduction that resembles the Josephson effect and the Coulomb blockade.

Below we set $e$, $\hbar$ and $k_B$ to unity, restoring dimensional units in final expressions when necessary.

We derive boundary conditions using the ideas of the scattering approach (for a review see Ref.~\onlinecite{Buttiker}) We describe a 1D conductor as a potential barrier at $|x|<l/2$ with a channel forming a quantum wire along the $x$-axis. The wire is attached to two symmetric 2D or 3D metallic leads. The region of 
conducting wire is considered as a scatterer. Since the results are very similar for both contacts we consider here, for brevity, the left lead only and the result for the right one will be given without derivation.

Without the loss of generality we assume that longitudinal (along the $x$-axis) and transverse motions are separable. The longitudinal motion in the leads is characterized by wave vectors $k$ and energy $\varepsilon_l = \frac{k^2}{2m}$, and transverse motion is described by energy $\varepsilon_n$, the total energy being $\varepsilon = \varepsilon_l + \varepsilon_n$, where $n$ is an index labeling transverse energies. We assume that electrons in the leads do not interact. Then we solve an equation of motion for electronic field operators in the leads using the continuity of both the field operators and their derivatives at $|x|=l/2$. This allows us to express the solution for the $n$-th transverse eigenstate in terms of the field operator $\hat \psi_b$ at the boundary
\begin{equation}
\hat \psi (x) = \hat\psi_b \cos k x + \frac{1}{k} \partial_x \hat\psi_b \sin k x .
\label{psi-operator}
\end{equation}
This expression contains both incident and outgoing waves. According to the causality principle, the incident wave $\hat \psi_{in}(x)$ is determined by a state of the lead far away from the barrier. Therefore, $\hat \psi_{in}(x)$ must not depend on properties of the barrier.
Equating the incoming part of Eq.~(\ref{psi-operator}) to the form describing free particles we find
\begin{equation}
\hat\psi_b  - \frac{i}{k_l} \partial_x \hat\psi_b = \frac{4\pi}{\sqrt{L}} \sum_{k>0} \hat c_{n,k} \delta (\varepsilon - \varepsilon_n - \frac{k^2}{2m}),
\label{psi-bc}
\end{equation}
where $k_l = \sqrt{2m(\varepsilon - \varepsilon_n)}$ and $\hat c_{n,k}$ is an annihilation operator of an electron in the lead with a longitudinal momentum $k$ in the $n$-th transverse mode of the lead.

Eq. (\ref{psi-bc}) relates the field operator at the boundary to the equilibrium states of the $n$-th transverse mode of energy $\varepsilon$ in the lead. We are interested in finding a relation between the boundary value of the field operator corresponding to the lowest transverse eigenstate of the conducting wire and the incident state of the lead. To find this relation, we project Eq.~(\ref{psi-bc}) onto the eigenstates of the wire. Since transverse states of the lead are not eigenstates of the wire, we obtain an infinite system of linear equations for boundary values of the field operators $\hat \psi_j$ of the transverse eigenstates $j$ of the wire
\begin{equation}
\hat\psi_j  - \sum_{j'} r_{jj'} \partial_x \hat\psi_{jj'} = \hat Z_j,
\label{psi-bc-system}
\end{equation}
We must find a solution of Eqs.~(\ref{psi-bc-system}) for the state $j=0$ describing the lowest subband which is responsible for an electronic transport in the wire, while the states $j>0$ with higher transverse energies do not contribute to the transport. It follows from  Eq.~(\ref{psi-bc-system}) that the relation we are looking for has a form
\begin{equation}
A(\varepsilon)\hat\psi_0  + B(\varepsilon) \partial_x \hat\psi_0 =  \frac{1}{\sqrt{V}} \sum_{\mathbf{n}=n,k>0} \gamma(k)\hat  c_{\mathbf{n}} 2\pi \delta(\varepsilon - \varepsilon_{\mathbf{n}}),
\label{psi-bc-0ch}
\end{equation}
where the exact expressions for the coefficients in Eq.~(\ref{psi-bc-0ch}) depend on the shape of the contacts. The boundary condition for the right contact has the same form but with complex-conjugate coefficients.

Coefficients in Eq.~(\ref{psi-bc-0ch}) are not arbitrary. In particular, they must provide correct anticommutation relations for electronic field operators. It is worthwhile to relate the coefficients to such physical parameters of the system as transmission probability $t$ of incident electrons. Therefore, we consider the system of noninteracting electrons, for which we can easily solve the equations for the field operators inside the wire. Then we impose a requirement of fulfillment of anticommutation relations, and calculate the conductance. This allows us to reduce the number of undetermined constants. As it is more convenient to express boundary conditions in terms of physical values, we multiply Eq.~(\ref{psi-bc-0ch}) on the left by its Hermitian conjugate, then we transform the obtained equation to the time representation assuming that the coefficients are slowly varying functions of energy in the region close to the Fermi energy. Finally, we find boundary condition for the left (right) contact
\begin{equation}
\frac{v_F}{t}\hat \rho \pm \hat j + v_F f \hat \rho_F  =
\frac{1}{V} \sum_{\mathbf{n},\mathbf{n'}}\hat c_{\mathbf{n'}}^+ \hat c_{\mathbf{n}} e^{i(\varepsilon_{\mathbf{n'}}-\varepsilon_{\mathbf{n}})t},
\label{bc-rho}
\end{equation}
where $\hat j$ and $\hat \rho$ are operators of current and of the smooth part of charge density  perturbations, $\hat \rho_F$ is the $2k_F$-component of charge density, which is related to the FO, $f$ is a number of
the order unity if the transmission probability is not close to unity, and $f \simeq \sqrt{2(1-t)}$ if $1-t\ll 1$. Thus the FOs disappear if the contacts are adiabatic.

In order to check the validity of conditions~(\ref{bc-rho}), we considered a wire with noninteracting 1D electrons with smoothly widening contacts, so that the contacts are nearly adiabatic. We also assumed that there might be a potential step of the height $U_0 \ll \varepsilon_F$ at the interface. Under these assumptions we were able to use the quasiclassical approximation in the lead and match the quasiclassical solution outside the 1D conductor with the exact solution inside the channel. We found that the condition~(\ref{bc-rho}) yielded a correct result for the conductance $G=tG_0$ in agreement with the Landauer formula. Here $G_0 = e^2/h$ is the conductance quantum.

In order to take into account interaction in the quantum wire, we consider spinless (spin-polarized) electrons described by a bosonic displacement field $\hat\Phi_\rho$ obeying the Tomonaga-Luttinger Hamiltonian~\cite{Giamarchi}. The bosonic field $\hat \Phi_\rho$ determines current and perturbations of charge density by means of relations
\begin{equation}
\hat \rho = -\frac{1}{\pi} \frac{\partial \hat\Phi}{\partial x} + \frac{k_F}{\pi }  \cos{ (2k_F x - 2 \hat\Phi)} ,\quad \hat I = (e /\pi)\partial_t \hat \Phi.
\label{i}
\end{equation}
The interaction is assumed to be short-range, described by the parameter $K_\rho \leq 1$ characterizing the strength of interaction ($K_\rho = 1$ for noninteracting electrons). A short-range interaction corresponds to gated quantum wires where the long-range part of interaction is screened by 3D gate electrodes. For quantum wires we can roughly estimate $K_\rho \sim \sqrt{\hbar v_F \epsilon/e^2}\approx $ 0.2 $\sqrt{\epsilon v_F\mbox{(cm/s)}/ 10^7},$ where $\epsilon$ is the background dielectric constant.

In the Heisenberg representation $\hat \Phi_\rho$ satisfies the wave equation~\cite{Giamarchi}
\begin{equation}
\left(  v^2 \partial^2_{x} - \partial^2_{t}\right)\hat\Phi_{\rho} (t,x) = 0
\label{phi}
\end{equation}
where $v = v_F / K_\rho$ is the velocity of plasma waves. This equation must be solved with boundary conditions at the contact which we obtain after bosonization of (\ref{bc-rho}). We should note that since one assumes a linear dispersion of electrons within the LL theory, the theory is valid provided that all energies are small in comparison with the Fermi energy. However, the term which is responsible for the FO after bosonization has the form $\frac{\sqrt{2(1-t)}}{\pi} \varepsilon_F \cos (2\hat\Phi_{\rho} + k_F l)$. Generally, this value is of order of the Fermi energy and it is small only if $\sqrt{1-t} \ll 1$. Therefore, we consider nearly adiabatic contacts, 
where $\sqrt{1-t} \ll 1$. Further, following Ref.~\cite{Grabert} we take into account screening of the potential of the leads by a 3D gate. Finally, we obtain the boundary conditions for bosonic field $\hat \Phi_\rho$ at the left and right contacts in the form
\begin{equation}
\frac{v_F}{K_\rho^2} \partial_x \hat\Phi_{\rho} \mp \partial_t \hat\Phi_{\rho} +  f \varepsilon_F \cos (2\hat\Phi_{\rho}  \mp k_Fl)  = \hat P_{L,R}.
\label{bc-operator}
\end{equation}
Now we represent the bosonic field operator as a sum of its expectation value and a fluctuating part, $\phi = \langle \hat\Phi_\rho\rangle$, $\hat\Phi_\rho (x=\mp l/2) = \phi_{L,R} + \hat\varphi_{L,R}$. Then we perform thermodynamic averaging of both sides of Eq.~(\ref{bc-operator}) and obtain boundary conditions for the expectation values
\begin{equation}
\frac{v_F}{K_\rho^2} \partial_x \phi_{L,R} \mp \partial_t \phi_{L,R} + d \sin (2\phi_{L,R} +\alpha)   = U_{L,R},
\label{bc-average}
\end{equation}
where $U_{L,R}$ is a potential applied to the left (right) contact, $d$ and $\alpha$ are given by expressions $d = f \varepsilon_F \sqrt{\langle \cos 2 \hat\varphi \rangle^2 + \langle \sin 2 \hat\varphi \rangle^2}$, $\alpha = \arctan \frac{\langle \cos 2 \hat\varphi \rangle }{ \langle \sin 2 \hat\varphi \rangle}$ for each contact. The equation for the expectation values $\phi_{L,R}$ which determines current and smooth perturbations of charge density contains the term related to the FO. This term depends on fluctuations in the 1D channel. If the contact is adiabatic $f=0$ this term disappears, and Eq.~(\ref{bc-average}) is reduced to boundary conditions by Egger and Grabert \cite{Grabert}.

In order to find correlation functions for the bosonic fields, we need to find first the correlation functions for fluctuating parts of the operators $\delta\hat P_{L,R} = \hat P_{L,R} - \langle \hat P_{L,R} \rangle$. In frequency representation it reads
\begin{equation}
\langle \{\delta\hat P_{L} (\omega),\delta\hat P_{L} (\omega')\} \rangle = 4\pi^2  \omega \coth\frac{\omega}{2T} \delta (\omega + \omega').
\label{P-corr-func}
\end{equation}
Now we can calculate the current induced by voltage $V=U_R-U_L$ applied to the leads. If the contacts are adiabatic then $f=\sqrt{2(1-t)}=0$, and there is no FO at the contacts. In this case, we obtain an ohmic current $j= \partial_t \Phi/\pi = G_0 V$. The result is different if the contacts are not adiabatic. In this case, we solve Eq.~(\ref{phi}) by performing the Fourier transformation and substituting then the solution into the boundary conditions given by Eq.~(\ref{bc-operator}). In this way we derive the equations for the field operators $\hat\Phi_{L,R}$ at the corresponding contacts.
\begin{eqnarray}
&&
A(\omega)\hat\Phi_{L}(\omega) + B(\omega)\hat\Phi_{R}(\omega) + \hat S_L(\omega) =\hat P_L(\omega), \nonumber \\
&&
B(\omega)\hat\Phi_{L}(\omega) + A(\omega)\hat\Phi_{R}(\omega) + \hat S_R(\omega) =\hat P_R(\omega),
\label{LR}
\end{eqnarray}
where $A(\omega) =\omega(i -\frac{1}{K_\rho \sin \omega t_l})$, $B(\omega) =\omega \cot \omega t_l$, $t_l = l/v$, $\hat S_{L,R}(\omega) = f \varepsilon_F \int dt e^{i\omega t}\cos (2\hat\Phi_{L,R} (t) \mp k_Fl)$. We cannot solve these nonlinear equations  easily, the main difficulty being the account of fluctuations. We assume that fluctuations are Gaussian. Strictly speaking, the fluctuations are not Gaussian. However, our approach can be justified strictly in the case of strong interelectronic repulsion and in the limit of high voltages, where the non-Gaussian part of fluctuations is small. This can be shown similarly to the case of a current passing through an impurity in a 1D conductor~\cite{ASVR}.

We consider several limiting cases which can be solved analytically. First, we try to find a stationary solution in the case of low applied potentials $\pm V/2$. After averaging Eqs.~(\ref{LR}), we obtain the following equation for each contact
\begin{equation}
d \sin (2\phi_{L,R} +\alpha) =  V/2.
\label{stat}
\end{equation}
Eq.~(\ref{stat}) has stationary solutions for a finite voltage when $d\neq 0$. We can calculate $d$ by using 
self-consistent harmonic approximation~\cite{Giamarchi}, in which fluctuations are assumed to be Gaussian. In this approximation, we replace $\sin 2 \hat\varphi$ with $e^{-2 \langle \hat\varphi^2 \rangle} 2 \hat\varphi$,  and obtain a simple expression
\begin{equation}
d= 2 f \varepsilon_F  e^{-2 \langle \hat\varphi^2 \rangle}  .
\label{scha}
\end{equation}
We have obtained linear equations for fluctuations which can be solved easily. Thus we can find $\hat \varphi_{L,R}$ and using the expression for anticommutators of fluctuation sources~(\ref{P-corr-func}) we can calculate the mean square of fluctuations
\begin{equation}
\langle \hat\varphi_{L,R}^2 \rangle = \int \frac{d\omega}{2\pi} \langle \hat\varphi_{L,R}^2(\omega) \rangle
\label{phi2}
\end{equation}
In pure Luttinger liquid without FOs at the contacts, this integral diverges logarithmically both at high and low frequencies. At high frequencies it must be cut off at the energies of the order of the Fermi energy. The divergence of fluctuations at low frequencies is a feature of one-dimensional systems. In our case we obtain that the integral is cut off at the lower limit of the order of $d$. As the complete expression for $\langle \hat\varphi_{L,R}^2(\omega) \rangle$ is rather cumbersome we give the result only for frequencies $\omega > d$ , which determine a large logarithmic contribution to $\langle \hat\varphi_{L,R}^2 \rangle$. Since the result is the same for both the contacts, we omit the indices $L,R$
\begin{equation}
\langle \hat\varphi^2(\omega) \rangle = \frac{\pi \coth \frac{\omega}{2T}}{\omega} \frac{K_\rho^2 [2 - (1-K_\rho^2)\sin^2 \omega t_l]}{4K_\rho^2  + (1-K_\rho^2)^2\sin^2 \omega t_l}.
\label{phi2ome}
\end{equation}
In addition to the logarithmically divergent part, this expression contains the oscillating factor induced by reflections of fluctuations from contacts. If the length of the quantum wire is large enough, $l \gg v/d $, these oscillations contribute little to the integral and the oscillating factor can be replaced with its average value $K_\rho/(1 + K_\rho)$. Further, the result of integration depends on a relation between $d$ and temperature $T$. At low temperatures $T\ll d$,  the integration yields with a logarithmic accuracy
\begin{equation}
\langle \hat\varphi^2 \rangle = \frac{K_\rho}{1 + K_\rho} \ln \frac{\varepsilon_F}{d \cos 2\phi}.
\label{phi2s}
\end{equation}
Since $d$ depends on $\langle \hat\varphi^2 \rangle$, Eq.~(\ref{phi2s}) is self-consistency condition for $\langle \hat\varphi^2 \rangle$. Substituting $d$ from Eq.~(\ref{scha}), we find $d$ and the maximal value of the left hand side of Eq.~(\ref{stat}). Thus we determine the value of the threshold voltage for which a static solution for mean phase $\phi$ exists.
\begin{equation}
V_T \simeq f^{\frac{1+ K_\rho}{1 - K_\rho}} \varepsilon_F .
\label{VT}
\end{equation}
We see that in the case of interelectronic repulsion, $K_\rho < 1$, the mean square of fluctuations $\langle \hat\varphi^2 \rangle$ and the amplitude of the FO are finite, while in the case of noninteracting system, when $K_\rho = 1$, fluctuations become infinite and $d=0$. Note that the role of the FOs here is similar to that of impurities in LL. Thus we obtain that if the voltage applied to the contacts is small enough, $V<V_T$, current does not flow through contacts. This result is a consequence of the approximation in which only  Gaussian fluctuations were taken into account. If we took into account fluctuations of solitonic type for which the phase increases by $2\pi$ due to tunneling, we would obtain a small tunneling current at $V<V_T$. Fluctuations of this type were studied in the case of impurities and they resulted in power-law I-V curves~\cite{Giamarchi}.

Now, we consider the case $T>d$. The self-consistency equation has solutions which correspond to a finite value of fluctuations only if $T < T_0 \sim f^{\frac{1 + K_\rho}{1 - K_\rho}} \varepsilon_F$, so at temperatures $T>T_0$ the FOs are destroyed by thermal fluctuations and do not affect electronic transport.
 
In the case of a short enough channel $l \ll v/d$ we must not average Eq.~(\ref{phi2ome}) over oscillations at $\omega t_l < 1$. At these frequencies $\langle \hat\varphi^2(\omega) \rangle$ in Eq.~(\ref{phi2ome}) is proportional to $\omega^{-1}$ as before but with a different factor. As a consequence the FOs do not affect the conduction when $l \ll v/V_T$.

If we increase the voltage above $V_T$, a transition to a nonstationary regime of conduction starts. We consider the limiting cases of low temperatures $T \ll V_T$ and of long wire $l \gg v/V_T$. It is difficult to obtain I-V curves at low voltages accurately, the main difficulty being the account of fluctuations with the mean square value periodically depending on time. The problem is simplified at high voltages $V \gg V_T$ when the mean square value becomes nearly constant with small oscillating component. In this case Eqs.~(\ref{LR}) can be solved perturbatively assuming that the oscillating part is small both in the fluctuations $\langle \hat\varphi^2 \rangle$ and in the mean phase $\langle \phi \rangle$. Note that Eqs.~(\ref{LR}) describe two interacting nonlinear oscillators. In such systems different solutions can exist. We consider a solution for which oscillations at both contacts are synchronous. Then the averaged Eqs.~(\ref{LR}) can be reduced to a single equation, which for the constant applied voltage reads
\begin{equation}
\partial_t \phi +   \int_0^\infty dt_1  Z(t-t_1) d(t_1) \sin 2 \phi (t_1) =  V/2.
\label{nonstat}
\end{equation}
To solve this equation we need to calculate the value of $d(t)$ which is determined by fluctuations. In order to do this we solve Eqs.~(\ref{LR}) for fluctuations substituting $d(t)$ in the form $d(t) = d_0 + d_c\cos \omega_0 t + d_s \sin \omega_0 t$, and assume that $d_c,d_s \ll d$. Thus we obtain a system of linear equations for $\hat \varphi_{L,R}$, from which linear equations for correlation functions $\langle \{  \hat\varphi_{R,L} (\omega),\hat\varphi_{R,L} (\omega')\} \rangle $ can be easily derived. After a rather cumbersome but not too difficult derivation we find that the main logarithmic contribution to $d$ is determined by Eq.~(\ref{phi2ome}) as before, but with different infrared cut-off frequency $b \sim d^2/\omega_0 \ll d$. Thus the self-consistency equation is again of the form~(\ref{phi2s}), but with $b$ instead of $d\cos 2\phi$. From the self-consistency condition we find
\begin{equation}
d_0 =  V_T \left(\frac{ V_T}{\omega_0}\right)^{\frac{2 K_\rho}{1 - 3 K_\rho}}, \quad  d_c \sim  \frac{d_0^2}{\omega_0},  \quad d_s \sim  \frac{d_0^2}{\omega_0} \ln \frac{\omega_0}{d}.
\label{dV}
\end{equation}
We see that at high voltages the solution with finite amplitude of the FOs exists only when $K_\rho < 1/3$, i.e. when interelectronic interaction is strong enough. The result differs from the stationary case where fluctuations do not destroy FOs at any repulsion strength $K_\rho < 1$. The result also differs from the case of impurity, where the critical value at high voltages is $K_\rho = 1/2$.

Now we can solve Eq.~(\ref{nonstat}) easily in the limit of high voltages $V \gg V_T$, and to calculate current using Eq.~(\ref{i}). The total current consists of dc part, $\bar I = V G_0 - I_{nl}$, and ac part, $I_{ac} \sin \omega_0 t$, which oscillates with frequency $\omega_0 = 2\pi \bar I/e\approx eV/\hbar$
\begin{equation}
I_{ac} \simeq  2\pi   G_0 d_0 K_\rho^2 /(K_\rho^2 + \tan^2 \frac{\omega_0 t_l}{2} ), \, I_{nl} \simeq 2\pi \left(\frac{V_T}{V}\right) I_{ac}.
\nonumber
\end{equation}
The oscillating factor in these expressions is due to reflections of generated current pulses from the contacts.

The work was supported by Russian Foundation for Basic Research. A part of the research was performed in the frame of the CNRS-RAS-RFBR Associated European Laboratory ``Physical properties
of coherent electronic states in condensed matter'' between Institut N\'eel, CNRS
and IRE RAS.


\begin{thebibliography}{19}

\bibitem{Giamarchi}T.\,Giamarchi, \textit{Quantum Physics in One Dimension},
(Clarendon Press, Oxford, 2003).

\bibitem{Auslaender}O.\,M.\,Auslaender, H.\,Steinberg, A.\,Yacoby et al,  Science {\bf 308}, 88 (2005)

\bibitem{carbon}H.\,Ishii,  H.\,Kataura, H.\,Shiozawa et al.
Nature {\bf 426}, 540 (2003).

\bibitem{ARS}S.\,N.\,Artemenko, S.\,V.\,Remizov, D.\,S.\,Shapiro,  Pis'ma v ZhETF {\bf 87}, 792 (2009) [JETP Lett. \textbf{87}, 692 (2009)].

\bibitem{Grabert}  R.\,Egger, H.\,Grabert, Phys. Rev. Lett. \textbf{80}, 2255 (1998);  H.\,Grabert,  \textit{Transport in Single Channel Quantum Wires}  in {\sl Exotic States in Quantum Nanostructures}, Ed. S.\,Sarkar (Kluwer, Dordrecht, 2002)

\bibitem{Buttiker}Ya.\,M.\,Blanter, M.\,B\"uttiker,  Phys. Rep. {\bf 336}, 1 (2000)

\bibitem{ASVR}S.\,N.\,Artemenko, D.\,S.\,Shapiro, R.\,R.\,Vakhitov, S.\,V.\,Remizov.  Journal of Physics: Conference Series {\bf 193}, 012119 (2009).

\end{thebibliography}
\end{document}